\documentclass[11pt]{article}
\usepackage{moriond}

\newcommand{\Photo}{}

\newcommand{\eq}[1]{\begin{equation} #1 \end{equation}}
\newcommand{\av}[1]{\langle #1 \rangle}

\begin{document}
\vspace*{2cm}
\title{$B\to K^*\ell\ell$: The New Frontier of New Physics searches in Flavor}

\author{ Sebastien Descotes-Genon$^a$, Tobias Hurth$^b$, Joaquim Matias$^c$, Javier Virto$^{c,**}$\let\thefootnote\relax\footnote{$^{**}$ Speaker} \\\vspace{0.3cm}}

\address{
$^{a}$ Laboratoire de Physique Th\'eorique, CNRS/Univ. Paris-Sud 11 (UMR 8627)\\ 91405 Orsay Cedex, France\\[2mm]
$^{b}$ PRISMA Cluster of Excellence \&	Institute for Physics (THEP)\\  Johannes Gutenberg University, D-55099 Mainz, Germany\\[2mm]
$^{c}$ Universitat Aut\`onoma de Barcelona, 08193 Bellaterra, Barcelona}

\maketitle
{\abstract{The exclusive decay mode $B\to K^*\ell\ell$ has become one of the key players in the search for New Physics in flavor. An increasing number of observables are being studied experimentally, and with increasing precision. Its theoretical description is well under control, and an interesting set of clean observables have been identified, reducing further the theoretical uncertainties. Model-independent analyses show that in the near future this mode will either reveal NP through a pattern of correlated deviations from the SM, or pose very stringent constraints on radiative and semileptonic operators.}}

\section{Theoretical and Experimental status of $B\to K^*\ell\ell$}

The exclusive $B\to K^*\ell\ell$  decay is today among the most promising $b\to s$ penguin modes due to recent theoretical and experimental developments. Pioneering experimental analyses have been performed at the $B$-factories and the Tevatron~\cite{Belle,BaBar,CDF}, providing measurements of the branching ratio and some rate asymmetries based on a few hundred events. However, it is LHCb that has opened the door to precision measurements of the full angular distribution~\cite{LHCb}.

The decay is mediated by an effective hamiltonian that can be split in a ``semileptonic'' and a ``hadronic'' part, ${\cal H}_{\rm eff}={\cal H}_{\rm eff}^{\rm sl}+{\cal H}_{\rm eff}^{\rm had}$. The semileptonic hamiltonian is composed by the electromagnetic operators ${\cal O}^{(\prime)}_{7\gamma}$ and the semileptonic operators 
${\cal O}^{(\prime)}_{9,10,S,P,T}$, all of which receive potential New Physics (NP) contributions. The corresponding amplitude factorizes trivially in the naive sense:
\eq{{\cal A}^{\rm sl}=\langle K^*\ell\ell|{\cal H}_{\rm eff}^{\rm sl}|B\rangle = \sum_i f_i(C_7,C_{7'},\cdots) F_i^{B\to K^*},}
where $f_i$ are short-distance functions and $F_i(q^2)$ are (seven) $B\to K^*$ form factors, with $q^2$ the momentum transfer to the lepton pair.

The hadronic hamiltonian is composed by four-quark (and chromomagnetic) operators and contributes to the semileptonic amplitude through the matrix element of the non-local operator $T\{j_{\rm em}^\mu(x) {\cal H}_{\rm eff}^{\rm had}(0)\}$, where $j_{\rm em}^\mu$ is an electromagnetic quark current:
\eq{{\cal A}^{\rm had}=i\frac{e^2}{q^2}\langle\ell^+\ell^- | \bar l \gamma_\mu l |0\rangle
\int d^4x\, e^{iq\cdot x}\langle K^*|T\{j_{\rm em}^\mu(x) {\cal H}_{\rm eff}^{\rm had}(0)\} |B\rangle.
}
This amplitude is non-factorizable in part, and it is not expected to receive any significant New Physics contributions due to the existing constraints from non-leptonic $B$ decays.

The theoretical difficulty with this exclusive decay is therefore two-fold: 1) determination of the hadronic form-factors, and 2) computation of the hadronic contributions.

A crucial step in the computation of the hadronic contributions consists in the identification of appropriate effective expansions in different regions of phase space. In the region of large recoil of the kaon ($E_{K^*}\gg \Lambda_{\rm QCD}$), an expansion in $\Lambda/m_b$ and $\Lambda/E_{K^*}$ can be performed. Up to perturbative and power corrections, all form factors can be expressed in terms of two ``soft'' form factors $\xi_{\|,\bot}$\cite{LEET}. One may also resort to QCDF/SCET to factorize the matrix elements \cite{BFS}. The hadronic amplitude may then be written as
\eq{{\cal A}^{\rm had}=C_a\xi_a + \Phi_B \otimes T_a \otimes \Phi_{K^*} + {\cal O}(\Lambda/m_b)}
where $C_a$, $T_a$ are perturbative quantities, $\xi_a$ are form factors and $\Phi_M$ are distribution amplitudes. The first term is of the same form as ${\cal A}^{\rm sl}$, while the second is not. This amplitude is known to leading order in $\Lambda/m_b$ and NLO in $\alpha_s$. 

In the region of low hadronic recoil, the momentum transfer $q^2$ --which corresponds to the invariant mass of the dilepton-- is large: $\sqrt{q^2}\gg E_{K^*},\Lambda$, and one can perform an OPE for the operator
${\cal K}_H^\mu(q)=\int d^4x\, e^{iq\cdot x} \,T\{j_{\rm em}^\mu(x) {\cal H}_{\rm eff}^{\rm had}(0)\}$~\cite{GP,BBF}:
\eq{{\cal K}_H^\mu(q)=\sum_i C_i(q) \,{\cal O}^\mu_i\ .}
Counting $q^2\sim m_b^2$ as same order in the power counting, the coefficients scale as $C_i \sim m_b^{3-d}$ where $d$ is the dimension of the corresponding operator ${\cal O}^\mu_i$; these operators start at dimension 3. The OPE might be performed within HQET~\cite{GP} or using the full QCD $b$-quark fields~\cite{BBF}, and it is known 
up to operators of dimension 5 (of order $(\Lambda/m_b)^2$). Within HQET several form factor relations arise, similarly to the case of large recoil \cite{GP,BHV}, which allows to build several ``clean observables''~\cite{BHV} in the low recoil region (see Section \ref{secclean}).

\begin{figure}
\centerline{
\includegraphics[width=6cm]{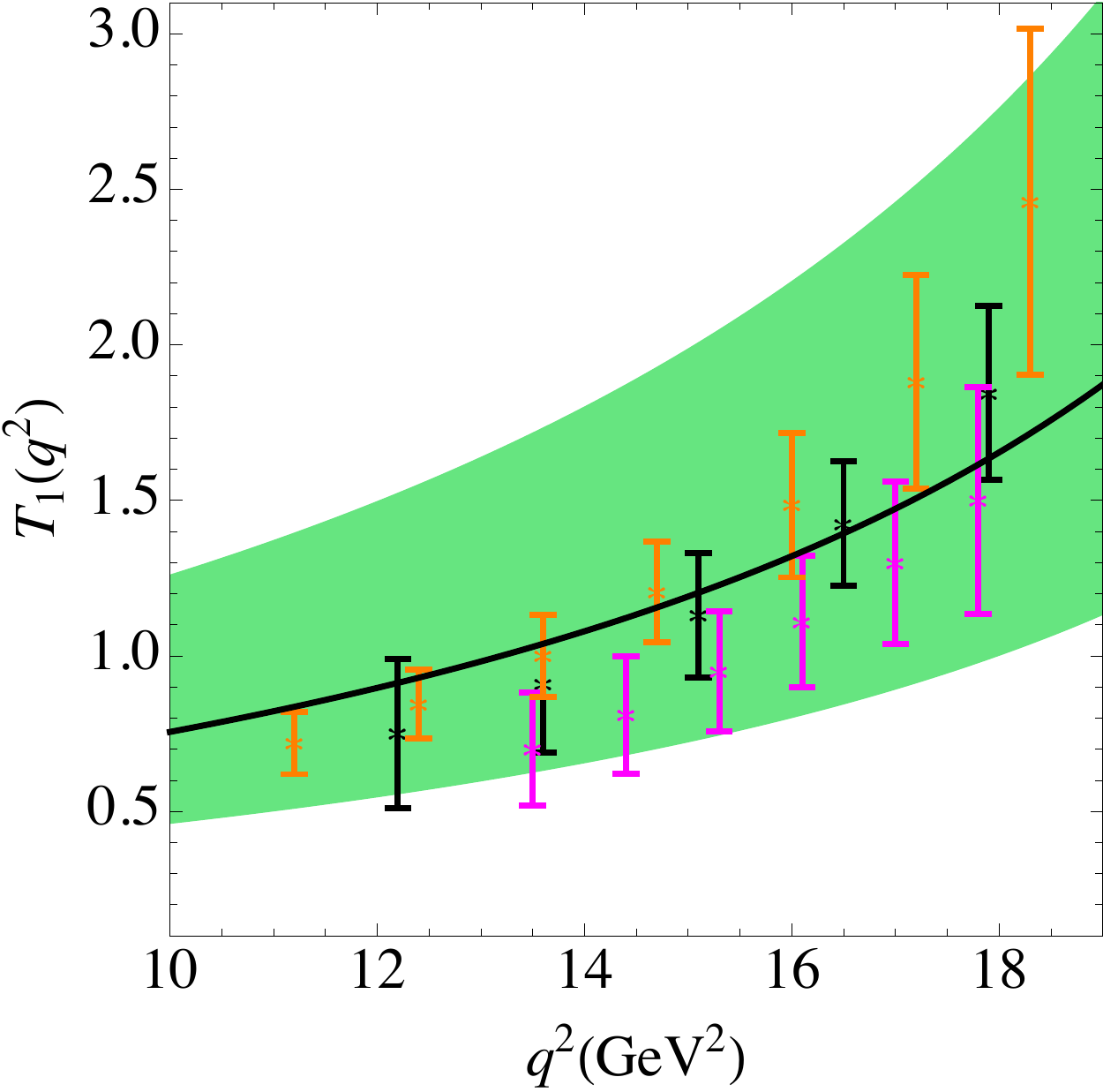}\hspace{1.5cm}
\includegraphics[width=6cm]{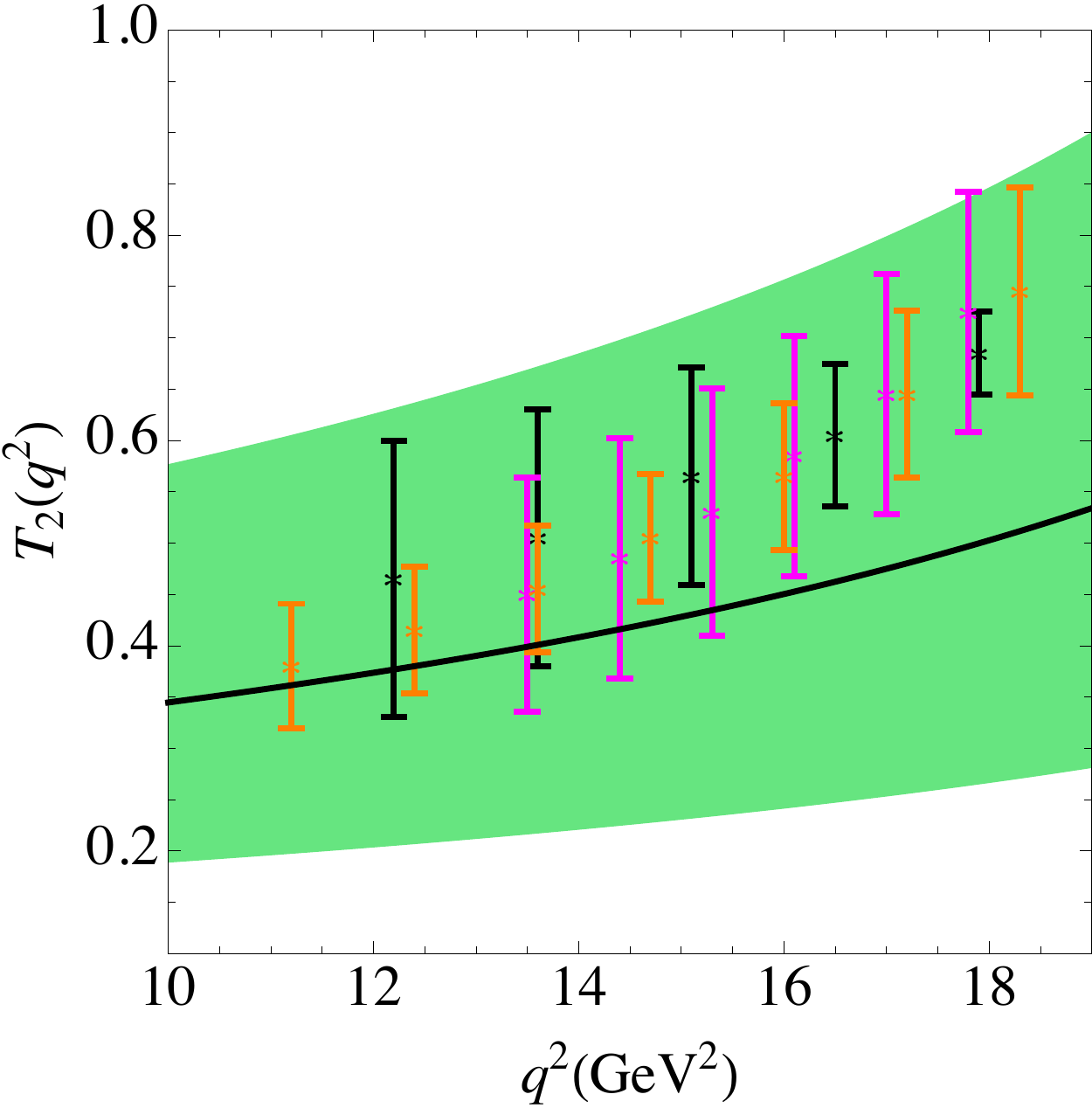}}
\caption{
Tensor form factors $T_1$ (left) and $T_2$ (right) at low recoil, obtained from $V$ and $A_1$ in Ref.~10 after imposing the HQET relations for $R_1=T_1/V$, $R_2=T_2/A_1$, and including a $20\%$ $\Lambda/m_b$ correction (green bands). The three sets of lattice data points correspond to the three sets of lattice QCD results in Table~1 of Ref.~12.}
\label{ffs}
\end{figure}

Both descriptions at low and large recoil are affected by potential corrections that are difficult to estimate. At large recoil, non-factorizable power corrections are unknown and could give non-negligible contributions to the amplitude as well as to the form factor relations. Long-distance effects from charm or light-quark loops could also play a role. Charm-loop effects are assumed to be more relevant since they are not suppressed by small CKM elements or small Wilson coefficients. These contributions have been studied recently~\cite{KMPW}, and it has been argued that their contribution to positive helicity amplitudes is suppressed and under control~\cite{CJ}, with the conclusion that several transverse asymmetries are genuine null tests of the SM. At low recoil, power corrections to form factor relations could also be an issue, as well as duality violations to the OPE. Model studies of duality violations suggest that these corrections should be small, specially when considering observables integrated over sufficiently broad $q^2$ bins~\cite{BBF}. 

Form factors have been studied in different contexts. Lattice QCD results are available for $T_1$ and $T_2$ at low recoil~\cite{Latt}. Light-cone sum rules have also been used to estimate the full set of form factors at maximum recoil, either with light-meson~\cite{BZ} or $B$-meson distribution amplitudes (DAs)~\cite{KMPW}, being the later method less precise due to our yet poor knowledge of the $B$-meson DA. Other possibilities that have been explored are the use of Dyson-Schwinger equations~\cite{dse}, or a fit to the data from short-distance-free asymmetries~\cite{HH}. The conclusion is that all determinations are compatible among each other if one is willing to accept conservative uncertainties. In Fig.~\ref{ffs} we show the compatibility between: (1) lattice results~\cite{Latt}, and (2) $T_1$, $T_2$ obtained from $V$ and $A_1$ as given in Ref.~\cite{KMPW}, after extrapolating to the low recoil region and using the HQET form factor relations with an arbitrary $20\%$ $\Lambda/m_b$ correction.

\section{Clean CP-averaged and CP-violating Observables}
\label{secclean}

In order to minimize the theoretical error associated to form factor uncertainties, one may study certain observables which have a reduced dependence on form factors. These observables are known as ``clean observables''. They have been constructed in the context of $B\to K^*\ell\ell$ for quite some time, leading to a large set of interesting observables with varying properties. A rather complete list of such observables is: $A_T^{\rm (i)}$~\cite{KM,EHMRR}, $A_T^{\rm (re,im)}$~\cite{BS}, $P_i$~\cite{MMRV}, $M_{1,2},S_{1,2}$~\cite{MMRV}, $P'_i$~\cite{DMRV} and $P_i^{\rm (\prime) CP}$~\cite{DHMV} at large recoil, and $H_T^{(i)}$, $a_{\rm CP}^{(i)}$~\cite{BHV} at low recoil.

The rational behind the construction of clean observables in both kinematic regions is the same. In both cases the form factor ratios $R_1=T_1/V$, $R_2=T_2/A$ and $R_3=T_{23}/A_{12}$ are predictions of the corresponding effective theories: $R_i=1$ up to perturbative and power corrections~\cite{DHMV}. This means that, up to such corrections and up to non-factorizable hadronic contributions to the amplitudes, the transversity amplitudes are proportional to a single form factor. Moreover, this form factor is the same for $L$ and $R$ amplitudes:
\eq{A_{\bot}^{L,R} = X_\bot^{L,R} V(q^2)\ ;\quad A_{\|}^{L,R} = X_\|^{L,R} A_1(q^2)\ ;\quad A_{0}^{L,R} = X_0^{L,R} A_{12}(q^2)\ ,}
where $X_{\bot,\|,0}$ are short distance functions. Clean observables are then constructed as ratios of amplitudes where the form factors cancel. All clean observables constructed in this way are clean in both kinematic regions, as for example $P_4=H_T^{(1)}$ or $P_5=H_T^{(2)}$. In addition, at large recoil the ratio $R_4=V/A_1$ is also predicted to be $1$ (but not at low recoil). Therefore some observables such as $P_1=A_T^{(2)}$ or $P_3$ are clean only at large recoil.

\begin{figure}
\centerline{
\includegraphics[width=3.8cm]{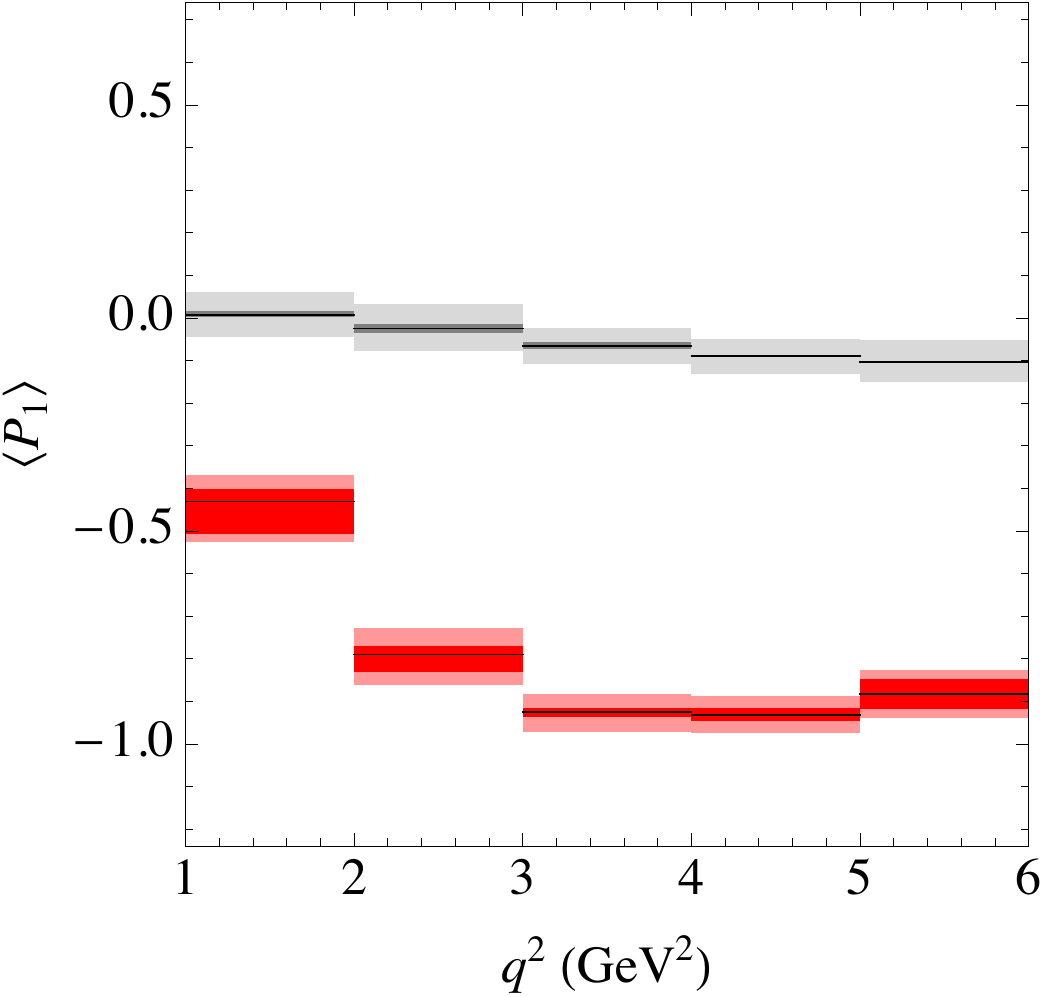}\hspace{0.1cm}
\includegraphics[width=3.8cm]{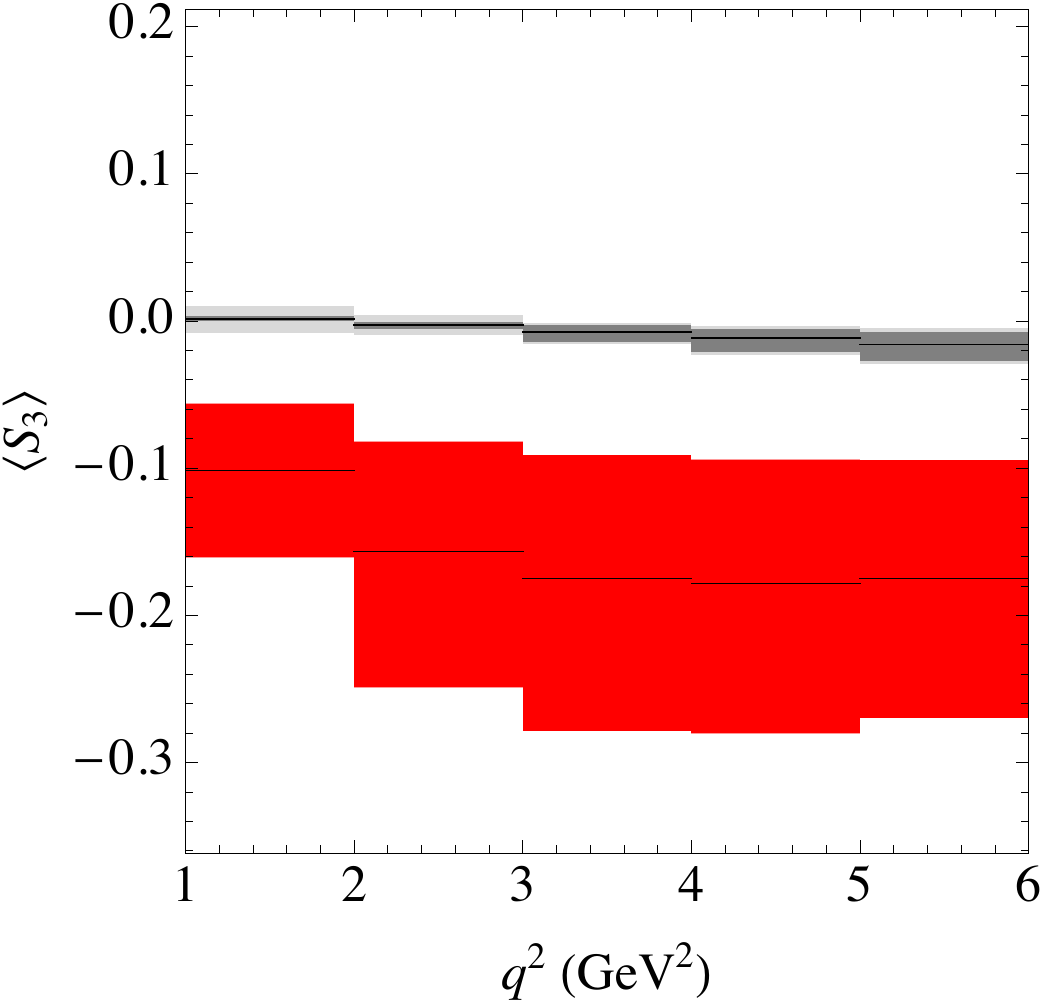}\hspace{0.1cm}
\includegraphics[width=3.9cm]{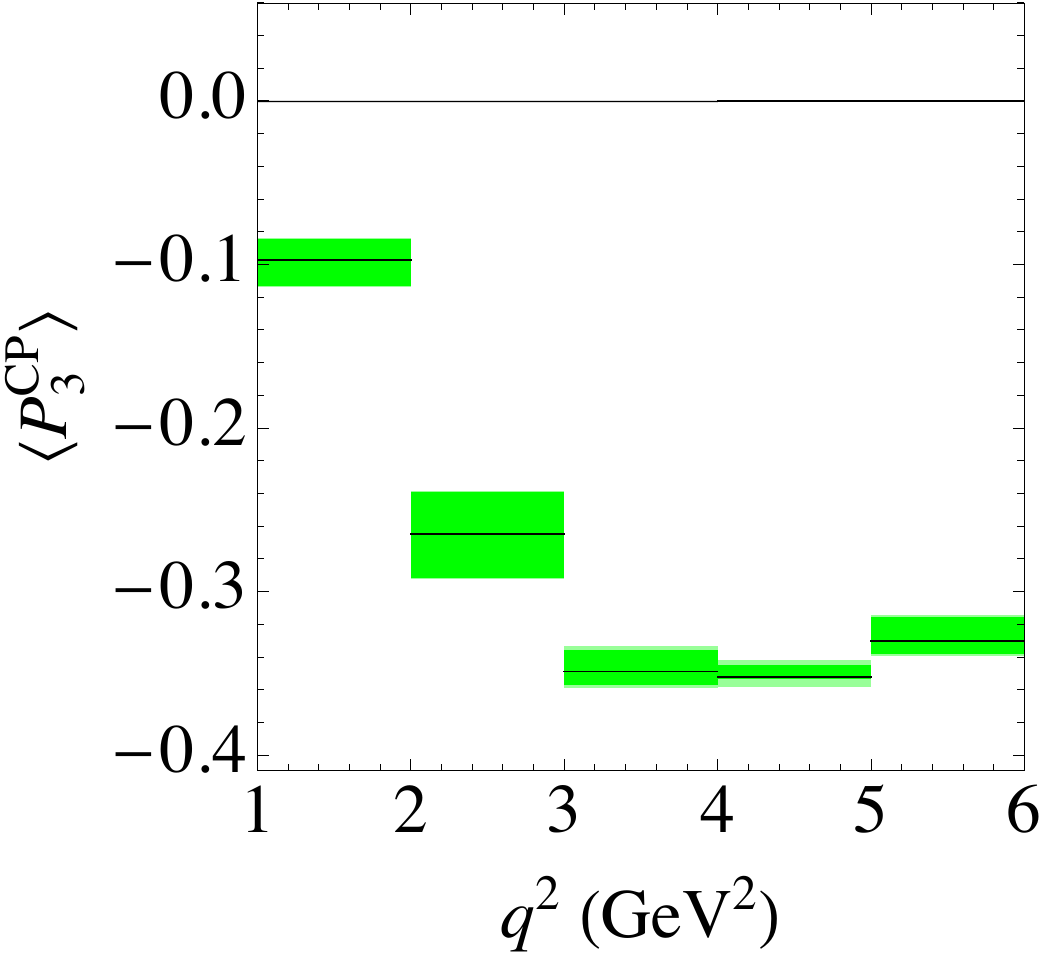}\hspace{0.1cm}
\includegraphics[width=3.9cm]{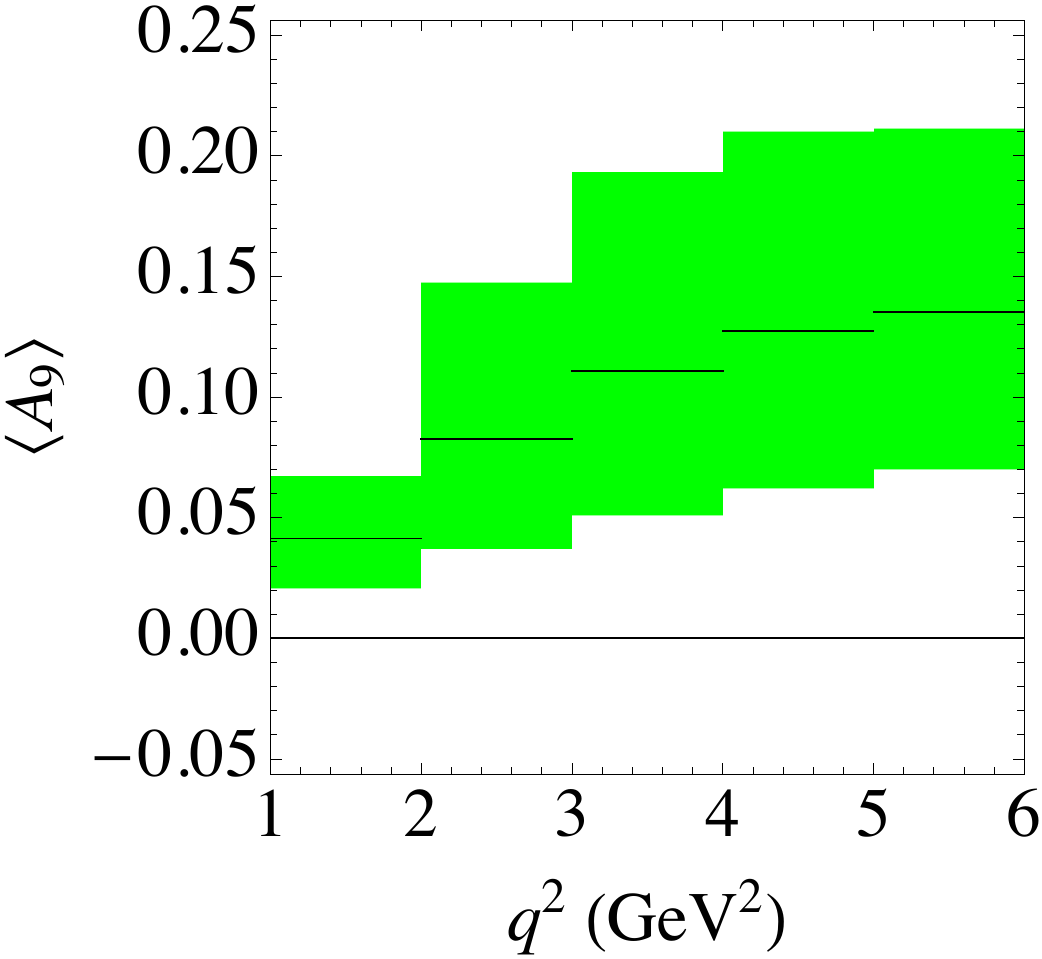}
}
\caption{$\av{P_1}$, $\av{S_3}$, $\av{P_3^{\rm CP}}$ and $\av{A_9}$ in the SM (gray) and in the case of a NP scenario consistent with all constraints from rare $B$ decays (red and green). In the presence of NP, the uncertainties for non-clean observables blow up, while clean observables remain clean. A $10\%$ estimate of power corrections is included in the uncertainties. 
}
\label{s3a9}
\end{figure}

Not all observables built in this way can be obtained \emph{exclusively} from the angular distribution, as transversity amplitudes contain also information on lepton polarizations. This will happen only if such observable can be written in terms of the coefficients of the distribution. In terms of amplitudes, a necessary condition is that the observable respects certain \emph{symmetries}, defined as ``transformations among the amplitudes that leave the angular distribution invariant''~\cite{EHMRR}. These symmetries are known explicitly in the massless case~\cite{EHMRR}, as well as in the case of massive leptons and in the presence of scalar operators~\cite{MMRV}.
As a byproduct of the symmetry formalism, the number of symmetries allows one to identify the number of experimental degrees of freedom in each case, pointing to a number of dependencies among the angular coefficients in the distribution. Even if in the most general cases all dependencies are lifted and all coefficients become formally independent (e.g. massive leptons plus scalar effects), it is clear that, since such effects are generally small, very acute correlations between certain coefficients will persist. These correlations should be handled with care. One possibility is to disentangle symmetry-preserving and symmetry-breaking effects already at the level of the observables. A set of observables that is always independent is composed by, for example, the following set of 8 observables:
~\cite{MMRV,DMRV}: $O_{m\ell=0}=\{d\Gamma/dq^2, F_L, P_1, P_2,P_3,P_4',P_5',P_6'\}$. One can then include four additional clean observables~\cite{MMRV}: $M_{1,2}$, $S_{1,2}$. $M_{1,2}$ are identically zero in the limit of massless leptons. They measure the breaking of two dependencies that arise in the massless limit. In the case of an experimental analysis that has not enough precision to resolve mass effects, these observables should be put to zero. Analogously, $S_{1,2}$ measure the  breaking of two dependencies that arise in the scalar-less limit, and should not be considered if scalar operators are much strongly constrained from elsewhere.

Clean observables have nonetheless a residual dependence on form factors. Perturbative symmetry-breaking corrections to form factor ratios constitute a (calculable) source of form factor dependence, as do also (incalculable) power corrections. Another source of form factor dependence comes from the fact that cancellations are achieved at each value of $q^2$, while real observables are measured in $q^2$ bins, meaning that theoretically the numerator and the denominator are integrated separately in $q^2$, and the leading-order cancellation is not exact. In practice, however, clean observables are seen to be quite clean even after including such calculable corrections and adding estimated power corrections. As an example, $\av{P_1}$ is noticeably independent of a variation of the uncertainties of the form factors, while $\av{F_L}$ is not~\cite{procF}.

\begin{figure}
\centerline{
\includegraphics[width=7cm]{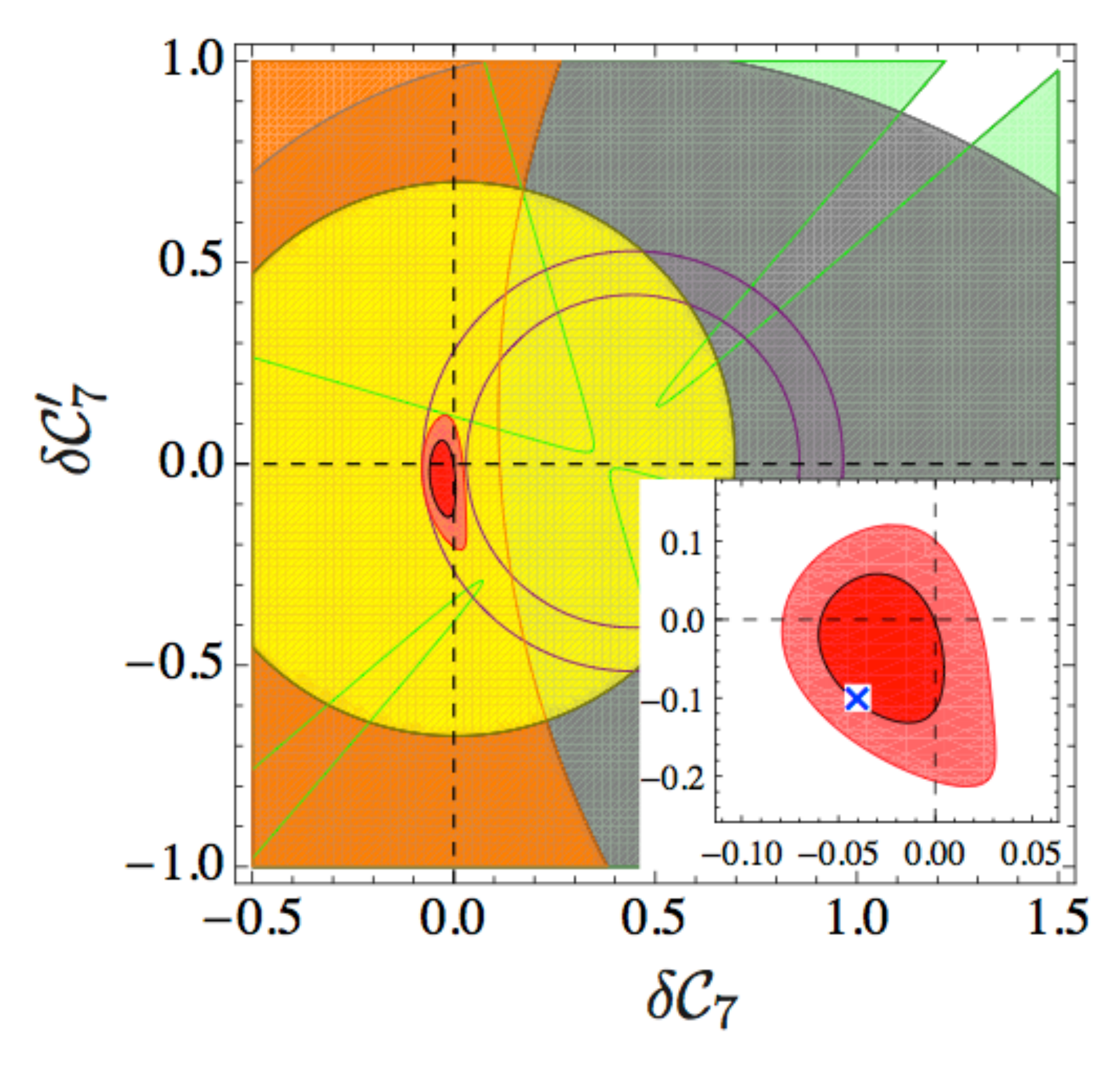}\hspace{1cm}
\includegraphics[width=7cm]{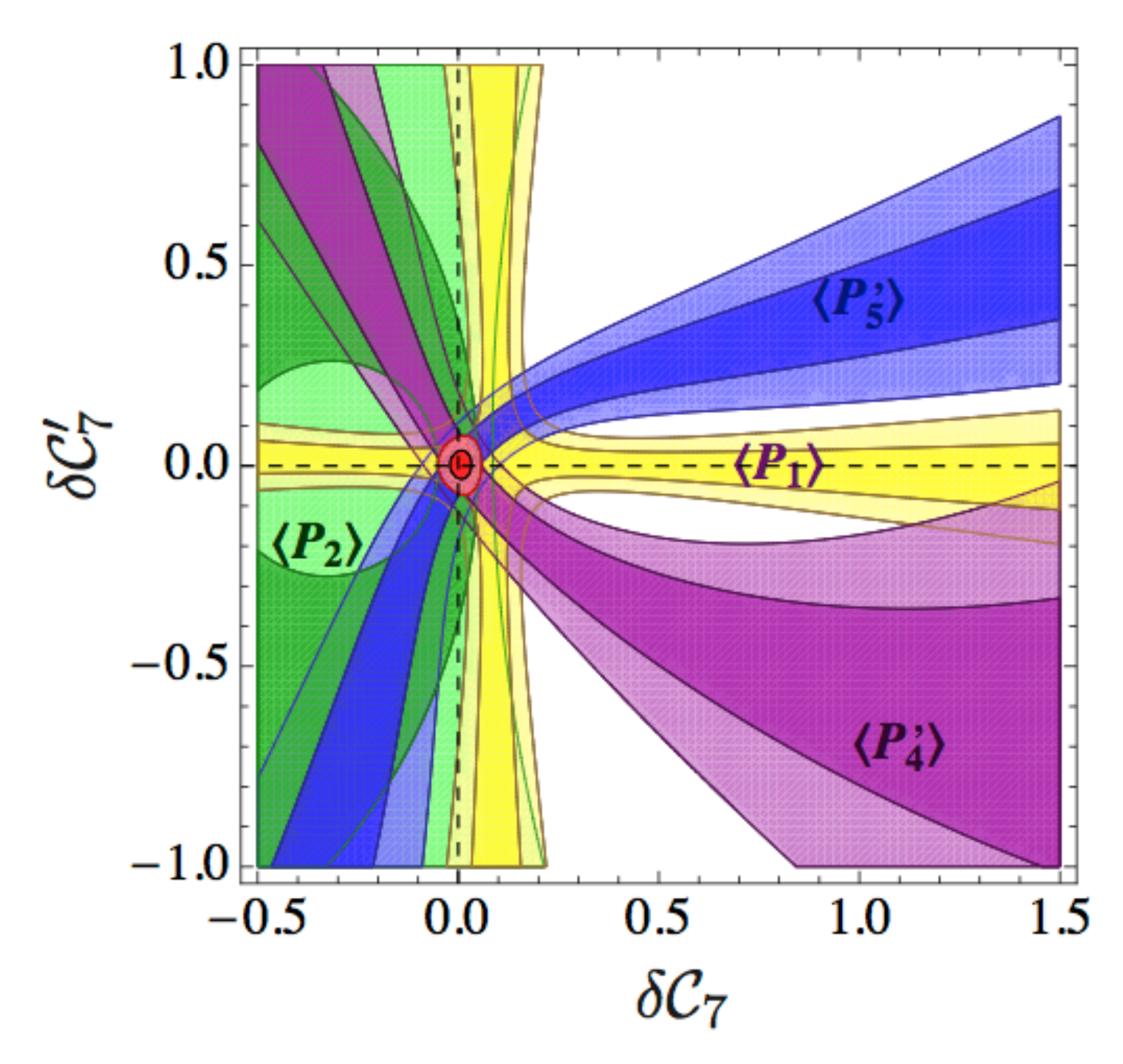}}
\caption{
Left: Constraints on the $C_7$-$C_7'$ plane from selected radiative and semileptonic decays (see the text). Right: Constraints on $C_7$-$C_7'$ from hypothetical future measurements of $\av{P_1}$, $\av{P_2}$, $\av{P_4}$, $\av{P_5}$ in the bin $[2.4.3]$, with central values equal to their SM prediction and uncertainties of $\sigma = 0.1$. Dark is 68.3\% and light 95.5\% C.L.}
\label{cons}
\end{figure}

There are some examples of observables that are not clean according to the definition adopted here, but that show very small uncertainties within the SM. For example, at large recoil,  $S_3$ and $A_9$~\cite{ABBBSW} are very close to zero in the SM and with small uncertainties. One might therefore wonder why to use the corresponding clean observables $P_1$ and $P_3^{\rm CP}$ instead. The coefficients $J_3$ and $J_9$ of the angular distribution are proportional to the helicity amplitudes $H_{V,A}^+$. In the SM (and a large recoil) these helicity amplitudes are very suppressed because the helicity form factors $V_+$, $T_+$ vanish, and the corresponding hadronic contribution is also suppressed. This is a very robust statement~\cite{CJ}. Since $S_3$, $P_1$ and $A_9$, $P_3^{\rm CP}$ are respectively proportional to $J_3$ and $J_9$, they are all genuine null tests of the SM. However this is no longer the case in the presence of nonzero contributions to $C_7'$, $C_9'$ or $C_{10}'$. This is shown in Fig.~\ref{s3a9}, where one can see how the uncertainties in $S_3$ and $A_9$ blow up in the presence of NP, while $P_1$ and $P_3^{\rm CP}$ are well behaved~\cite{DHMV}. It is therefore very important to focus on clean observables in NP studies.

\section{Model-Independent Constraints and prospects for $B\to K^*\mu^+\mu^-$}

Rare $B$ decays already constrain quite strongly the possible NP contributions to radiative and semileptonic operators, being the magnetic operators $C_7^{(')}$ the most constrained. In this sense, the branching ratio $BR(B\to X_s\gamma)$, the CP asymmetry $S_{K^*\gamma}$ and the isospin asymmetry $A_I(B\to X_s\gamma)$ reduce very significantly the allowed space in the $C_7$-$C_7'$ plane. The semileptonic observables $BR(B\to X_s\mu^+\mu^-)$, $\av{A_{\rm FB}(B\to K^*\mu^+\mu^-)}$ and $\av{F_L(B\to K^*\mu^+\mu^-)}$ help in excluding isolated regions away from the SM point (e.g. the ``flipped sign'' solution for $C_7$), but only if we ban NP contributions to semileptonic operators (see Fig.~\ref{cons}). Within this scenario, the prospects from clean observables in $B\to K^*\mu^+\mu^-$ are excellent, since a measurement of $P_{1,2}$ and $P'_{4,5}$ in the second experimental bin with a precision of $\sigma\sim 0.1$ would alone already provide a tighter constraint than the other radiative and semileptonic modes together~\cite{DMRV} (see Fig.~\ref{cons}).  Constraints in more general scenarios when some or all semileptonic operators receive NP contributions can be found in Ref.~\cite{DMRV}. In this cases, constraints from $B\to K\mu^+\mu^-$ and $B_s\to \mu^+\mu^-$ can also be relevant. The later has been extensively discussed after its first experimental evidence \cite{LHCbBsmumu}, and the recent SM reevaluation~\cite{THBsmumu}, which takes into account the finite $B_s$ width difference relevant for branching ratio measurements at LHCb~\cite{deltaGs}.

\begin{figure}
\centerline{
\includegraphics[width=16cm]{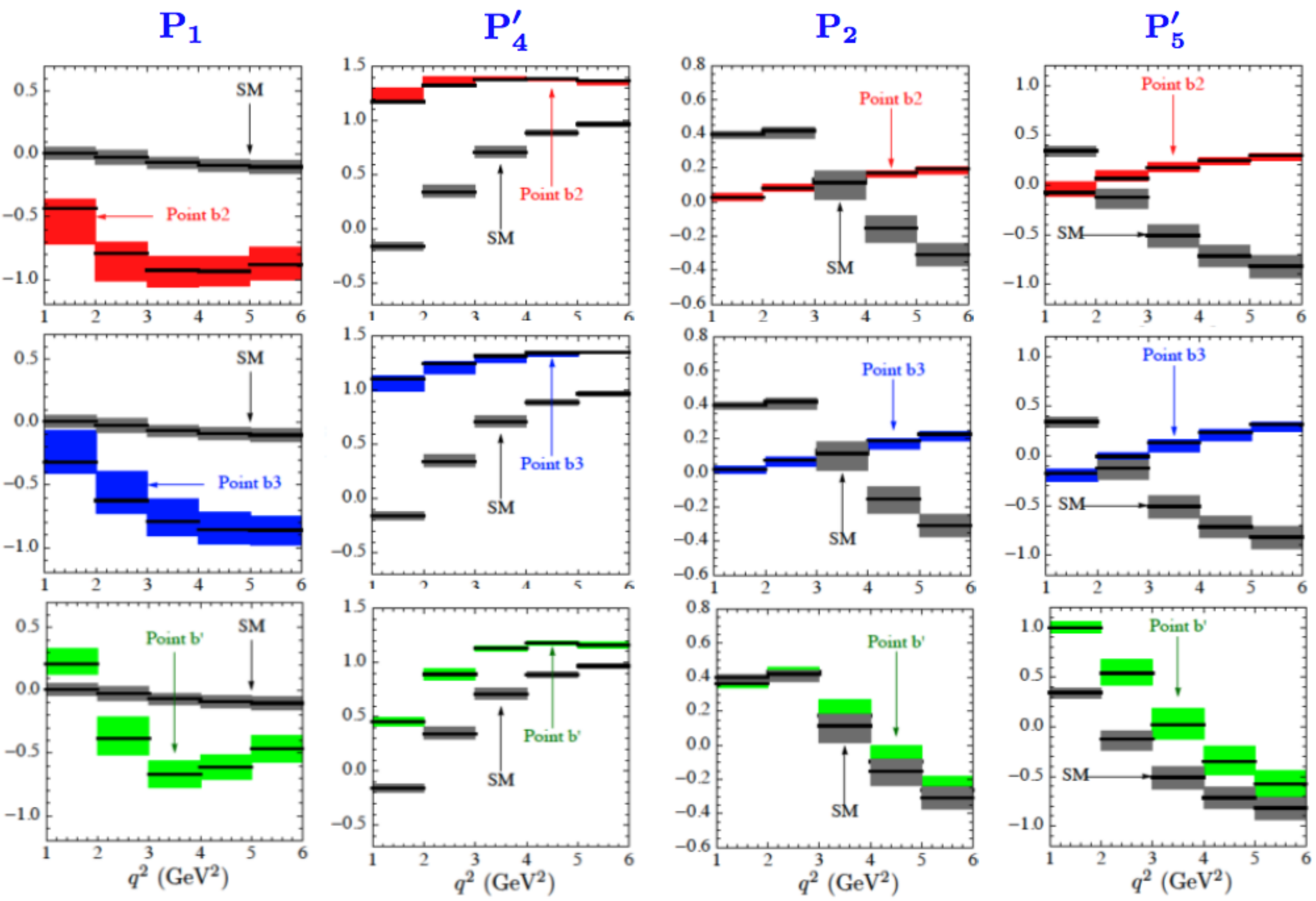}
}
\caption{
New Physics complementarity of  the clean observables $\av{P_1}$, $\av{P_2}$, $\av{P_4'}$, $\av{P_5'}$, exemplified through the NP benchmark points $b2$, $b3$ and $b'$ defined in Ref.~20, which are NP points compatible with all existing bounds.}
\label{comp}
\end{figure}

These constraints allow to identify different benchmark NP scenarios consistent with current bounds that can be used to test the future opportunities of $B\to K^*\mu^+\mu^-$. Complementarity is a crucial feature in this context, since a NP discovery in flavor physics must be backed up by a full set of correlated deviations from the SM in well theoretically controlled observables. An example of such New Physics complementarity is shown in Fig.~\ref{comp}. One can see that considering different observables in different bins is key not only to discover NP, but also to identify it.

\section{Conclusions}

Among the large set of rare $b\to s$ penguin modes, the exclusive decay $B\to K^*\ell\ell$ has attracted a lot of attention recently. This attention is well justified. Experimental analyses are accessing more and more angular observables, with increasing precision, and the prospects are bright. On the theory side, complete sets of CP-averaged and CP-violating clean observables in both kinematic regions of interest are known, and the understanding of the relevant uncertainties is in good shape. Model-independent analyses show that in the near future $B\to K^*\mu^+\mu^-$ will either reveal for the first time an important set of correlated deviations from the SM, or pose the most restrictive set of constraints on radiative and semileptonic operators.

\begin{figure}
\includegraphics[width=16.5cm]{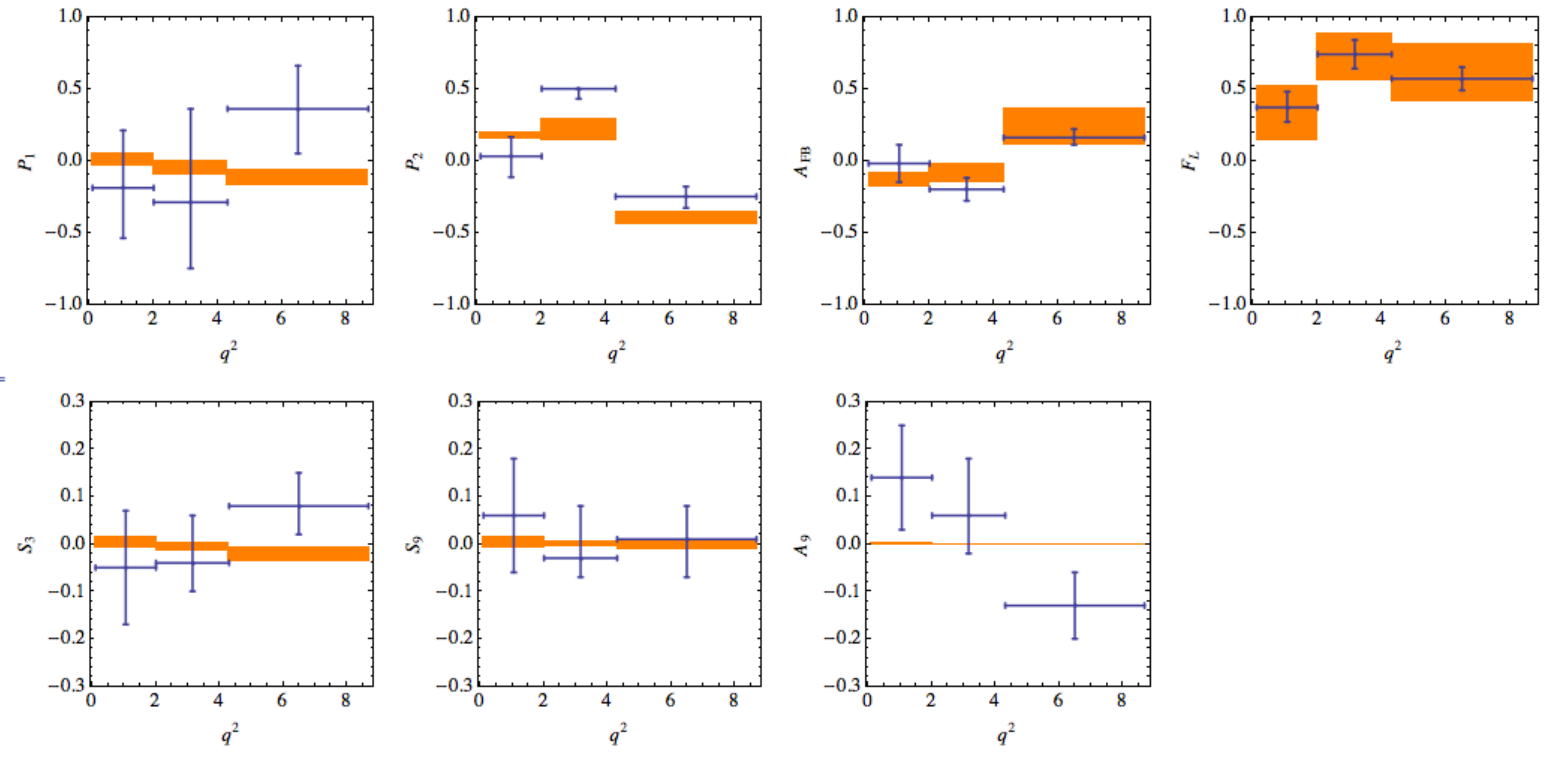}
\caption{
Latest results from LHCb on the angular distribution of $B\to K^*\mu^+\mu^-$ (blue), compared to the theoretical predictions (orange) from Ref.~21. }
\label{new}
\end{figure}

Latest experimental news are even more exciting. LHCb has reported a measurement of the observables $P_1$, $P_2$, $A_{\rm FB}$, $F_L$, $S_3$, $S_9$ and $A_9$ obtained only from the first 1 fb$^{-1}$ collected at the detector~\cite{LHCbnew}. These results, although still with large uncertainties in some cases, already pose interesting constraints on NP. In Fig.~\ref{new} we show the experimental results for the three bins in the large recoil region, compared to our SM predictions~\cite{DHMV}. Future analyses with the full data set collected in the first run of the LHC are certainly much awaited and will have surely a significant impact on flavor physics.

\section*{Acknowledgments}

It is a pleasure to thank the organizers of the EW session of Rencontres de Moriond 2013 for the arrangement of a very stimulating workshop. J.V. is supported in part by ICREA-Academia funds and FPA2011-25948. J.M. enjoys financial support from grants FPA2011-25948 and SGR2009-00894.

\end{document}